\documentclass{article}

\PassOptionsToPackage{authoryear, round}{natbib}

\usepackage[dblblindworkshop, final]{neurips_2025}
\workshoptitle{The 4th Deep Learning for Code Workshop}

\usepackage[utf8]{inputenc} 
\usepackage[T1]{fontenc}    
\usepackage{hyperref}       
\usepackage{url}            
\usepackage{booktabs}       
\usepackage{amsfonts}       
\usepackage{nicefrac}       
\usepackage{microtype}      
\usepackage{xcolor}         

\usepackage{listings}

\definecolor{lightgray}{gray}{0.97}

\usepackage{pifont}
\newcommand{\cmark}{\ding{51}} 
\newcommand{\xmark}{\ding{55}} 

\usepackage{enumitem}

\usepackage{titlesec}
\titlespacing*{\paragraph}
  {0pt}   
  {0pt}   
  {1em}   

\title{Agentic Property-Based Testing:\\Finding Bugs Across the Python Ecosystem}

\author{%
  Muhammad Maaz \\
  MATS, Anthropic \\
  \And
  Liam DeVoe \\
  Northeastern University \\
  \And
  Zac Hatfield-Dodds \\
  Anthropic \\
  \And
  Nicholas Carlini \\
  Anthropic
}

\begin{document}

\maketitle

\begin{abstract}
Property-based testing (PBT) is a lightweight formal method, typically implemented as a randomized testing framework. Users specify the input domain for their test using combinators supplied by the PBT framework, and the expected properties or invariants as a unit-test function. The framework then searches for a counterexample, e.g. by generating inputs and calling the test function. In this work, we demonstrate an LLM-based agent which analyzes Python modules, infers function-specific and cross-function properties from code and documentation, synthesizes and executes PBTs, reflects on outputs of these tests to confirm true bugs, and finally outputs actionable bug reports for the developer. We perform an extensive evaluation of our agent across 100 popular Python packages. Of the bug reports generated by the agent, we found after manual review that 56\% were valid bugs and 32\% were valid bugs that we would report to maintainers. We then developed a ranking rubric to surface high-priority valid bugs to developers, and found that of the 21 top-scoring bugs, 86\% were valid and 81\% we would report. The bugs span diverse failure modes from serialization failures to numerical precision errors to flawed cache implementations. We reported 5 bugs, 4 with patches, including to NumPy and cloud computing SDKs, with 3 patches merged successfully. Our results suggest that LLMs with PBT provides a rigorous and scalable method for autonomously testing software. 
Our code and artifacts are available at: \url{https://github.com/mmaaz-git/agentic-pbt}.
\end{abstract}

\section{Introduction}

Property-based testing (PBT) is a software testing paradigm that aims to verify whether a general property holds over a predefined input domain. In contrast to traditional example-based testing, it does not require specific examples of expected outputs. Instead, PBT allows the developer to define invariants that should hold for all inputs, which are then checked on a diverse set of automatically-generated inputs. A property might require that the output of a function is always non-negative, that some $f$ and $g$ commute for all $x$ ($f(g(x)) = g(f(x))$), or that an interpreted program has equivalent semantics to its compiled form. Popularized by QuickCheck in Haskell \citep{claessen2000quickcheck}, there now exist PBT libraries for many of the common programming languages, including Python's \textit{Hypothesis} \citep{maciver2019hypothesis}, the focus in this work. PBT can be more robust than example-based testing which inherently relies on the developer to anticipate edge cases \citep{maciver2019praise}.

Despite its theoretical appeal, property-based testing is less popular than example-based testing due to the challenge of identifying meaningful properties to test \citep{goldstein2024property}. Specifying meaningful properties, especially as codebases get more complex, often requires significant domain expertise and time investment. Recent advances in large language models (LLMs) have demonstrated remarkable code understanding, which makes it possible to \textit{automatically} mine for properties. However, existing work on using LLMs for PBTs focuses on a single function at a time, and has only a single generation step.

In our work, we introduce \textit{agentic property-based testing}, an approach leveraging the multi-step reasoning capabilities of coding agents and the rigor of PBT. We develop an agent that can autonomously crawl through and understand entire codebases, look for high-value properties (including across functions and classes), write PBTs and run them, and then analyze failing tests for validity. After the agent runs for several rounds, it either surfaces a bug, or states it is unable to identify one.

We demonstrate the effectiveness of agentic PBT through a large-scale evaluation across the Python ecosystem, testing 100 popular Python packages with billions of downloads. Manual evaluation of the bug reports generated  by our agent showed that it successfully identified genuine bugs in several widely-used libraries: for example, our agent  autonomously finds a bug in NumPy, the pre-eminent numerical library in Python with more than 500 million monthly downloads, which was acknowledged by the maintainers, showing the real-world applicability of our approach. This work represents the largest systematic evaluation of AI-driven property-based testing to date, establishing a new paradigm for scalable software auditing.

\section{Agent}

Our agent is built on top of Anthropic's Claude Code, a terminal-based coding agent \citep{anthropic2025claudecode}, that allows Claude to  execute bash commands and read/write files. Our agent is implemented as a natural language prompt, stored in a Markdown file, that is passed to Claude Code. As it is a prompt, our agent is easily portable to other agent implementations, like OpenAI's Codex \citet{openai_codex}, or general frameworks like ReAct \citep{yao2022react}. Our main contribution is our comprehensive evaluation which demonstrates that LLM-based PBT works at scale, finding diverse bugs with few false alarms.

Our agent is built to work with Python codebases, and generates  Hypothesis PBTs to test that code. It takes a single argument, which points the agent towards a particular target, either a single Python file (e.g., \texttt{normalizers.py}), module (e.g., \texttt{numpy}, \texttt{scipy.signal}), or function (e.g., \texttt{requests.get()}, \texttt{json.loads()}).

We developed the prompt by using industry best-practices \citep{anthropic2025bestpractices}, collating high-value information from the Hypothesis documentation \citep{hypothesis-docs}, and iterative refinement from manual inspection of agent runs. The full prompt is in Appendix \ref{sec:appendix-prompt}.

\paragraph{Prompt Overview}

The agent is given the following six instructions, paraphrased below:
\begin{enumerate}[noitemsep, topsep=0pt]
    \item Analyze the target: Figure out if the target is a module, function, or a file.
    \item Understand the target: Find and read documentation, function signatures, source code, etc.
    \item Propose properties: Look for properties grounded in the information from the past step. Some examples of high-value properties, e.g., invariants, round-trip, metamorphic, are given. Thoroughly understand the input domain, e.g., by looking at functions that call the function under test.
    \item Write tests: Translate these properties into Hypothesis property-based tests.
    \item Execute and triage tests: Run tests with \texttt{pytest}. For failing tests, reflect using a rubric. If the failing test is a false alarm, go to Step 4 and refine the testing strategy. For passing tests, verify that the test is meaningful.
    \item Report bugs: If convinced a bug is genuine, create a bug report in a Markdown file, following a standard format including a summary of the bug, the PBT that exposed the bug, a short reproducing script, why it is a bug, and possibly a proposed patch.
\end{enumerate}

Finally, we provide the agent with a short reference to Hypothesis, which shows essential patterns, some key testing principles, and links to online documentation which can be fetched if needed.

\paragraph{Key Design Choices}

Our agent maintains a to-do list to track its progress through the 6-step cycle, as well as to keep track of, e.g., which properties it would like to test. We attempt to reduce false alarms throughout by emphasizing that proposed properties should be strongly supported by evidence,  asking the agent to reflect on failing tests carefully, and to focus on a few high-value properties. Lastly, we let the agent have broad autonomy: it decides which functions to target, and decides when a failing property is truly a bug (with some direction from a rubric).

\section{Experiment}

We evaluate our agents ability to find bugs by curating a diverse corpus of $100$ Python packages. We combine three complementary sampling approaches to ensure comprehensive coverage:
\begin{itemize}[noitemsep]
    \item \textit{Hand-selected from the standard library ($n=15$):} \texttt{json}, \texttt{pathlib}, \texttt{collections}, \texttt{itertools}, \texttt{functools}, \texttt{datetime}, \texttt{re}, \texttt{os}, \texttt{urllib}, \texttt{statistics}, \texttt{decimal}, \texttt{base64}, \texttt{uuid}, \texttt{html}, \texttt{csv}.
    
    \item \textit{Hand-selected from third-party PyPI packages ($n=15$):} \texttt{numpy}, \texttt{pandas}, \texttt{requests}, \texttt{flask}, \texttt{sqlalchemy}, \texttt{matplotlib}, \texttt{scipy}, \texttt{beautifulsoup4}, \texttt{pydantic}, \texttt{fastapi}, \texttt{django}, \texttt{tornado}, \texttt{keras}, \texttt{httpie}, \texttt{black}.
    
    \item \textit{Random sample from the top 5,000 PyPI packages, by downloads ($n=70$):} Variety of domains including data, cloud computing, and machine learning. The full list is provided with our code.
\end{itemize}

For each package, we test the agent on the main module (e.g., for \texttt{beautifulsoup4}, the main module is \texttt{bs4}), and all submodules one level deep (e.g., \texttt{numpy.linalg}). In total, we ran on 933 modules.

We used Claude Opus 4.1 \cite{opus41} as the LLM for all agent runs. Each agent was run in an isolated virtual environment containing the package under test and its dependencies, along with Hypothesis. Agents were run in parallel, with up to $N=20$ concurrently. Agents had read/write permission within their working directory, could execute \texttt{python} and \texttt{pytest} bash commands, had read access to the virtual environment for source code, and internet access. The experiment was run on a RunPod Ubuntu 20.04 container with 8 vCPUs, 16 GB RAM, and 50 GB disk.

\paragraph{Evaluation} Definitive validation of reported bugs across such diverse libraries is difficult as it requires domain expertise in each specific codebase. First, we developed an initial scoring rubric with the goal of eliminating clear false alarms and got Claude Opus 4.1 to grade all reports. We randomly sampled $n=50$ bug reports from the top 80\% by score. Two of the authors independently scored each bug report, using two criteria: ``Is this a valid bug? If yes, would we reasonably report this to the maintainers?". Disagreements were discussed to arrive at a final consensus. Using insights from the initial rubric, we developed a final rubric (see our code) which scores bug reports out of 15, got Claude Opus 4.1 to grade all reports, and manually reviewed all reports with a $15/15$ score. To validate our agents with maintainers with actual domain expertise, we (manually) reported some particularly interesting bugs to their respective repositories.

\section{Results}

\paragraph{Summary statistics} We evaluated our agent across 100 Python packages covering 933 modules. The agent generated 984 bug reports, discovering issues in 786 modules (84.2\%) and averaging just over one bug per module. Total agent runtime was 136.6 hours (82 min/package, 8.8 min/module) and total API cost was \$5,474.20 (\$54.74/package, \$5.87/module, \$5.56/bug report). Individual module costs ranged from \$0.65 to \$15.42, with a median of \$5.81. Each agent run involved roughly 110 turns. In aggregate, the agents used 2.21 billion (input \& output) tokens (2.37 million/module).

\paragraph{Manual review} Inter-rater agreement on the initial review was $\kappa = 0.31$. After reaching consensus, 56.0\% (95\% CI: 42.2\%, 69.8\%) of reports were determined to be valid bugs, and 32.0\% (95\% CI: 19.1\%, 44.9\%) were valid and worth reporting. With the final rubric, $18$ of the $21$ top-scoring reports were valid, and $17$ of those were both valid and worth reporting.

\paragraph{Reported bugs}

While examining reports, we selected the following particularly interesting bugs to report to the original developer. The corresponding bug reports generated by the agent are in Appendix \ref{sec:bug_reports}. They demonstrate the range of issues discoverable through our approach:

\begin{description}[noitemsep, topsep=0pt]

\item \textbf{[numpy] (numpy.random)} \\
\textit{Property:} Wald distribution should only return non-negative samples. \\
\textit{Bug:} Negative samples sometimes returned due to catastrophic cancellation. \\
\textit{Status:} Bug acknowledged. Patch merged. \href{https://github.com/numpy/numpy/pull/29609}{[PR \#29609]}

\item \textbf{[aws-lambda-powertools] (aws\_lambda\_powertools.shared.functions)} \\
\textit{Property:} \texttt{slice\_dictionary()} splits a dictionary into chunks which should be able to be recombined into the original dictionary. \\
\textit{Bug:} It returns the same (first) chunk repeatedly, due to not incrementing the iterator. \\
\textit{Status:} Bug acknowledged. Patch merged. \href{https://github.com/aws-powertools/powertools-lambda-python/pull/7246}{[PR \#7246]}

\item \textbf{[cloudformation-cli-java-plugin] (rpdk.core.jsonutils.utils)} \\
\textit{Property:} \texttt{item\_hash()} function should produce different outputs for different inputs. \\
\textit{Bug:} All list inputs hash to $hash(\texttt{None})$, due to use of the in-place \texttt{.sort()} method, which returns \texttt{None}. \\
\textit{Status:} Patch submitted. \href{https://github.com/aws-cloudformation/cloudformation-cli/pull/1106}{[PR \#1106]}

\item \textbf{[tokenizers] (tokenizers.tools)} \\
\textit{Property:} \texttt{EncodingVisualizer.calculate\_label\_colors()} should return a valid HSL color format. \\
\textit{Bug:} The returned string is missing a closing parenthesis. \\
\textit{Status:} Bug acknowledged. Patch merged. \href{https://github.com/huggingface/tokenizers/pull/1853}{[PR \#1853]}

\item \textbf{[python-dateutil] (dateutil)} \\
\textit{Property:} \texttt{easter()} should be on a Sunday. \\
\textit{Bug:} Returns a non-Sunday date for the Julian calendar in various years. \\
\textit{Status:} Issue reported. Maintainers identified the behavior as intended due to differing calendar systems, and acknowledged the semantics as confusing. \href{https://github.com/dateutil/dateutil/issues/1437}{[Issue \#1437]}

\end{description}

\section{Discussion and Conclusion}

Our evaluation demonstrates that LLM-guided property-based testing can systematically uncover bugs missed by traditional testing. With a cost of \$5.56/bug report, and extrapolating from our manual grading that 56\% of these are valid bugs, our agent can find bugs for \$9.93/valid bug. This is an upper bound on the real-world cost, where developers with domain expertise can be more judicious with where to target the agent. The diversity of issues, spanning numerical issues to business logic issues, show the power of PBT, and the ability of agents to autonomously mine for such properties. 

\paragraph{Limitations} The primary limitation is that we did not manually review all 984 bug reports. However, our review of a subset of reports shows that the false discovery rate has a 95\% CI of [30.2\%, 57.8\%]. The secondary limitation is intent ambiguity: the agent cannot distinguish intentional design violations from bugs. An example of this is a bug report about \texttt{LookupDict} from the \texttt{requests} library: based on its name and the fact it subclasses from the built-in \texttt{dict}, our agent tested whether it exhibits \texttt{dict}-like behavior, which it in fact does not. The same issue was raised on GitHub in 2022\footnote{\url{https://github.com/psf/requests/issues/6238}}, but the maintainer replied that it was not meant to work like \texttt{dict}. In practice, both limitations can be mitigated by developers' domain expertise rather than blind exhaustive testing as we did. 

\paragraph{Comparison to Related Work}

\citet{vikram2023can} study converting library documentation into Hypothesis PBTs for 40 functions across 10 popular Python libraries (e.g., \texttt{numpy}). With their best model and prompting strategy, only 41\% of generated PBTs ran without error and passed on the implemented code, and at most 21\% of documented properties were captured. In contrast, we do better by using an agentic approach, which is better able to capture more complex properties due to multiple steps. More recent work \cite{he2025use} propose a different take on this problem: a ``generator" LLM produces candidate code while a ``tester" LLM synthesizes PBTs from the problem description. PBT failures are fed back to guide code refinement, yielding improvements over example-based test-driven development. There is also extensive work on LLMs for software testing in general: see \citet{wang2024software} for a review.

\paragraph{Conclusion} We presented an automated approach for bug discovery using LLM-guided property-based testing, which discovered real bugs across several popular Python libraries. The properties and failing test cases discovered show the utility of combining LLMs with PBT. While such a tool is useful for developers, a possible malicious use is autonomous discovery of vulnerabilities. As LLMs improve and get cheaper, agentic PBT could become an increasingly valuable complement to traditional testing, helping developers systematically explore code behavior and find bugs.

\newpage
\bibliographystyle{abbrvnat}
\bibliography{ref}

\begin{thebibliography}{13}
\providecommand{\natexlab}[1]{#1}
\providecommand{\url}[1]{\texttt{#1}}
\expandafter\ifx\csname urlstyle\endcsname\relax
  \providecommand{\doi}[1]{doi: #1}\else
  \providecommand{\doi}{doi: \begingroup \urlstyle{rm}\Url}\fi

\bibitem[{Anthropic}(2025{\natexlab{a}})]{anthropic2025bestpractices}
{Anthropic}.
\newblock {Claude Code: Best practices for agentic coding}.
\newblock \url{https://www.anthropic.com/engineering/claude-code-best-practices}, Apr. 2025{\natexlab{a}}.
\newblock Accessed: 2025-08-20.

\bibitem[{Anthropic}(2025{\natexlab{b}})]{anthropic2025claudecode}
{Anthropic}.
\newblock {Claude Code}.
\newblock \url{https://www.anthropic.com/claude-code}, 2025{\natexlab{b}}.
\newblock Accessed: 2025-08-20.

\bibitem[{Anthropic}(2025{\natexlab{c}})]{opus41}
{Anthropic}.
\newblock {System Card Addendum: Claude Opus 4.1}.
\newblock Technical report, Anthropic, Aug 2025{\natexlab{c}}.
\newblock URL \url{https://assets.anthropic.com/m/4c024b86c698d3d4/original/Claude-4-1-System-Card.pdf}.

\bibitem[Claessen and Hughes(2000)]{claessen2000quickcheck}
K.~Claessen and J.~Hughes.
\newblock {QuickCheck}: a lightweight tool for random testing of haskell programs.
\newblock In \emph{Proceedings of the fifth ACM SIGPLAN international conference on Functional programming}, pages 268--279, 2000.

\bibitem[Goldstein et~al.(2024)Goldstein, Cutler, Dickstein, Pierce, and Head]{goldstein2024property}
H.~Goldstein, J.~W. Cutler, D.~Dickstein, B.~C. Pierce, and A.~Head.
\newblock Property-based testing in practice.
\newblock In \emph{Proceedings of the IEEE/ACM 46th International Conference on Software Engineering}, pages 1--13, 2024.

\bibitem[He et~al.(2025)He, Chen, Zhang, Shao, Gao, and Sheng]{he2025use}
L.~He, Z.~Chen, Z.~Zhang, J.~Shao, X.~Gao, and L.~Sheng.
\newblock Use property-based testing to bridge llm code generation and validation.
\newblock \emph{arXiv preprint arXiv:2506.18315}, 2025.

\bibitem[{Hypothesis Contributors}(2025)]{hypothesis-docs}
{Hypothesis Contributors}.
\newblock Hypothesis documentation.
\newblock \url{https://hypothesis.readthedocs.io/en/latest/}, 2025.
\newblock Accessed: 2025-08-20.

\bibitem[MacIver(2019)]{maciver2019praise}
D.~MacIver.
\newblock In praise of property-based testing.
\newblock \emph{URL: https://increment. com/testing/in-praise-of-property-based-testing}, 2019.

\bibitem[MacIver et~al.(2019)MacIver, Hatfield-Dodds, et~al.]{maciver2019hypothesis}
D.~R. MacIver, Z.~Hatfield-Dodds, et~al.
\newblock Hypothesis: A new approach to property-based testing.
\newblock \emph{Journal of Open Source Software}, 4\penalty0 (43):\penalty0 1891, 2019.

\bibitem[OpenAI(2025)]{openai_codex}
OpenAI.
\newblock {OpenAI Codex}.
\newblock \url{https://openai.com/codex/}, 2025.
\newblock Accessed: 2025-08-20.

\bibitem[Vikram et~al.(2023)Vikram, Lemieux, Sunshine, and Padhye]{vikram2023can}
V.~Vikram, C.~Lemieux, J.~Sunshine, and R.~Padhye.
\newblock Can large language models write good property-based tests?
\newblock \emph{arXiv preprint arXiv:2307.04346}, 2023.

\bibitem[Wang et~al.(2024)Wang, Huang, Chen, Liu, Wang, and Wang]{wang2024software}
J.~Wang, Y.~Huang, C.~Chen, Z.~Liu, S.~Wang, and Q.~Wang.
\newblock Software testing with large language models: Survey, landscape, and vision.
\newblock \emph{IEEE Transactions on Software Engineering}, 50\penalty0 (4):\penalty0 911--936, 2024.

\bibitem[Yao et~al.(2022)Yao, Zhao, Yu, Du, Shafran, Narasimhan, and Cao]{yao2022react}
S.~Yao, J.~Zhao, D.~Yu, N.~Du, I.~Shafran, K.~Narasimhan, and Y.~Cao.
\newblock {ReAct}: Synergizing reasoning and acting in language models.
\newblock \emph{arXiv preprint arXiv:2210.03629}, 2022.
\newblock URL \url{https://arxiv.org/abs/2210.03629}.
\newblock Accessed: 2025-08-20.

\end{thebibliography}


\newpage
\appendix

\section{Agent prompt}
\label{sec:appendix-prompt}

Note: some characters may not render properly in the PDF; we provide the full prompt file in our code release.

\lstinputlisting[backgroundcolor=\color{lightgray}]{hypo.md}

\newpage
\section{Selected bug reports}
\label{sec:bug_reports}

A selection of bug reports, as written by the agent.

\subsection{numpy}
\begin{lstlisting}
# Bug Report: numpy.random.wald Produces Negative Values

**Target**: `numpy.random.wald`
**Severity**: High
**Bug Type**: Logic
**Date**: 2025-08-18

## Summary

The `numpy.random.wald` function produces negative values when called with large mean parameters (>= 1e8), violating the mathematical definition of the Wald (inverse Gaussian) distribution which only produces positive values.

## Property-Based Test

```python
import numpy.random
from hypothesis import given, strategies as st, settings

@given(
    st.floats(min_value=1e8, max_value=1e15),
    st.floats(min_value=0.1, max_value=10.0)
)
@settings(max_examples=50)
def test_wald_negative_values(mean, scale):
    """Wald distribution should never produce negative values"""
    samples = numpy.random.wald(mean, scale, size=1000)
    assert all(s >= 0 for s in samples), f"Found negative values with mean={mean}, scale={scale}"
```

**Failing input**: `mean=100000000.0, scale=1.099609375`

## Reproducing the Bug

```python
import numpy.random

numpy.random.seed(42)
mean = 1e8
scale = 1.0
samples = numpy.random.wald(mean, scale, size=1000)

negative_samples = samples[samples < 0]
print(f"Generated {len(samples)} samples")
print(f"Found {len(negative_samples)} negative values")
print(f"Minimum value: {samples.min()}")
```

## Why This Is A Bug

The Wald (inverse Gaussian) distribution is mathematically defined to only produce positive values. The probability density function is only defined for x > 0, and the documentation states that the function "Draw[s] samples from a Wald, or inverse Gaussian, distribution" with "mean : float or array_like of floats - Distribution mean, must be > 0".

Negative values violate this fundamental property and can cause downstream issues in applications that rely on the mathematical properties of the distribution, such as reliability modeling, financial modeling, and statistical inference.

## Fix

The issue likely stems from numerical precision problems when computing the inverse transform or rejection sampling for large mean values. The implementation should include bounds checking to ensure all generated values are strictly positive, possibly by:

1. Adding explicit bounds checking after sample generation
2. Using a more numerically stable algorithm for large mean values
3. Switching to an alternative sampling method when mean exceeds a certain threshold

```diff
# Conceptual fix in the sampling routine
def wald_sample(mean, scale):
    # ... existing sampling logic ...
    sample = compute_sample(mean, scale)
+   # Ensure non-negativity due to numerical issues
+   if sample < 0:
+       sample = 0.0  # or resample
    return sample
```
\end{lstlisting}

\subsection{aws-lambda-powertools}
\begin{lstlisting}
# Bug Report: aws_lambda_powertools.shared.functions.slice_dictionary Returns Duplicate Chunks

**Target**: `aws_lambda_powertools.shared.functions.slice_dictionary`
**Severity**: High
**Bug Type**: Logic
**Date**: 2025-08-18

## Summary

The `slice_dictionary` function incorrectly returns the same chunk multiple times instead of properly slicing the dictionary into different parts, causing data loss and duplication.

## Property-Based Test

```python
from hypothesis import given, strategies as st
from aws_lambda_powertools.shared.functions import slice_dictionary

@given(
    data=st.dictionaries(
        keys=st.text(min_size=1, max_size=20),
        values=st.one_of(st.integers(), st.text(), st.booleans()),
        min_size=0,
        max_size=100
    ),
    chunk_size=st.integers(min_value=1, max_value=20)
)
def test_slice_dictionary_confluence(data, chunk_size):
    """Test that sliced dictionary chunks can be reconstructed to the original."""
    chunks = list(slice_dictionary(data, chunk_size))

    # Reconstruct the dictionary from chunks
    reconstructed = {}
    for chunk in chunks:
        reconstructed.update(chunk)

    assert reconstructed == data, f"Reconstruction failed: {data} != {reconstructed}"
```

**Failing input**: `data={'0': 0, '00': 0}, chunk_size=1`

## Reproducing the Bug

```python
import sys
sys.path.insert(0, '/root/hypothesis-llm/envs/aws-lambda-powertools_env/lib/python3.13/site-packages')
from aws_lambda_powertools.shared.functions import slice_dictionary

# Example 1: Lost keys
data = {'0': 0, '00': 0}
chunks = list(slice_dictionary(data, chunk_size=1))
print(f"Original: {data}")
print(f"Chunks: {chunks}")
# Output: Chunks: [{'0': 0}, {'0': 0}]
# Expected: [{'0': 0}, {'00': 0}]

# Example 2: All chunks are identical
data = {'a': 1, 'b': 2, 'c': 3, 'd': 4, 'e': 5}
chunks = list(slice_dictionary(data, chunk_size=2))
print(f"Original: {data}")
print(f"Chunks: {chunks}")
# Output: [{'a': 1, 'b': 2}, {'a': 1, 'b': 2}, {'a': 1, 'b': 2}]
# Expected: [{'a': 1, 'b': 2}, {'c': 3, 'd': 4}, {'e': 5}]
```

## Why This Is A Bug

The function is supposed to split a dictionary into chunks of the specified size, but instead it repeatedly yields the same first `chunk_size` items. This violates the expected behavior of:
1. Each key appearing exactly once across all chunks
2. Chunks being different slices of the original dictionary
3. Being able to reconstruct the original dictionary from the chunks

## Fix

```diff
--- a/aws_lambda_powertools/shared/functions.py
+++ b/aws_lambda_powertools/shared/functions.py
@@ -141,5 +141,8 @@ def powertools_debug_is_set() -> bool:


 def slice_dictionary(data: dict, chunk_size: int) -> Generator[dict, None, None]:
-    for _ in range(0, len(data), chunk_size):
-        yield {dict_key: data[dict_key] for dict_key in itertools.islice(data, chunk_size)}
+    it = iter(data)
+    for _ in range(0, len(data), chunk_size):
+        chunk_keys = list(itertools.islice(it, chunk_size))
+        if chunk_keys:
+            yield {dict_key: data[dict_key] for dict_key in chunk_keys}
```
\end{lstlisting}

\subsection{cloudformation-cli-java-plugin}
\begin{lstlisting}
# Bug Report: rpdk.core.jsonutils.utils.item_hash All Lists Hash to Same Value

**Target**: `rpdk.core.jsonutils.utils.item_hash`
**Severity**: High
**Bug Type**: Logic
**Date**: 2025-08-18

## Summary

The `item_hash` function in `rpdk.core.jsonutils.utils` contains a critical bug that causes all list inputs to hash to the same value (the MD5 hash of "null"), completely breaking the hash function's purpose for list-type data.

## Property-Based Test

```python
from hypothesis import given, strategies as st
from rpdk.core.jsonutils.utils import item_hash

@given(st.lists(st.integers()), st.lists(st.integers()))
def test_item_hash_different_lists_different_hashes(list1, list2):
    """Different lists should produce different hashes (except in rare collisions)."""
    if list1 != list2:
        hash1 = item_hash(list1)
        hash2 = item_hash(list2)
        # This test fails because all lists hash to the same value
        assert hash1 != hash2 or list1 == list2
```

**Failing input**: Any two different lists, e.g., `[1, 2, 3]` and `[4, 5, 6]`

## Reproducing the Bug

```python
import sys
sys.path.insert(0, '/root/hypothesis-llm/envs/cloudformation-cli-java-plugin_env/lib/python3.13/site-packages')

from rpdk.core.jsonutils.utils import item_hash

list1 = [1, 2, 3]
list2 = [4, 5, 6]
list3 = ["a", "b", "c"]
empty_list = []

hash1 = item_hash(list1)
hash2 = item_hash(list2)
hash3 = item_hash(list3)
hash4 = item_hash(empty_list)

print(f"item_hash([1, 2, 3]) = {hash1}")
print(f"item_hash([4, 5, 6]) = {hash2}")
print(f"item_hash(['a', 'b', 'c']) = {hash3}")
print(f"item_hash([]) = {hash4}")

assert hash1 == hash2 == hash3 == hash4 == "37a6259cc0c1dae299a7866489dff0bd"
print("BUG: All lists hash to the same value!")
```

## Why This Is A Bug

The `item_hash` function is supposed to generate unique hashes for different inputs. However, due to a coding error on line 32, all list inputs produce the same hash value. This completely defeats the purpose of a hash function, which should produce different outputs for different inputs (except for rare collisions). The bug causes:

1. **Loss of uniqueness**: All lists, regardless of content, produce identical hash values
2. **Hash collisions**: Any code using this for deduplication or caching will fail
3. **Security implications**: If used for any security-sensitive hashing, this would be a critical vulnerability

## Fix

```diff
--- a/rpdk/core/jsonutils/utils.py
+++ b/rpdk/core/jsonutils/utils.py
@@ -29,7 +29,8 @@ def item_hash(
     if isinstance(item, dict):
         item = {k: item_hash(v) for k, v in item.items()}
     if isinstance(item, list):
-        item = [item_hash(i) for i in item].sort()
+        hashed_items = [item_hash(i) for i in item]
+        item = sorted(hashed_items)
     encoded = json.dumps(item, sort_keys=True).encode()
     dhash.update(encoded)
     return dhash.hexdigest()
```

The bug occurs because `.sort()` returns `None`, not the sorted list. The fix uses `sorted()` which returns the sorted list, or assigns the sorted list after calling `.sort()` on it.
\end{lstlisting}

\subsection{tokenizers}
\begin{lstlisting}
# Bug Report: tokenizers.tools.visualizer Missing Closing Parenthesis in HSL Color Format

**Target**: `tokenizers.tools.visualizer.EncodingVisualizer.calculate_label_colors`
**Severity**: Medium
**Bug Type**: Contract
**Date**: 2025-08-18

## Summary

The `calculate_label_colors` method generates malformed HSL color strings missing a closing parenthesis, causing CSS parsing errors when used in HTML visualization.

## Property-Based Test

```python
from hypothesis import given, strategies as st
import re
from tokenizers.tools import Annotation, EncodingVisualizer

@given(st.lists(st.text(min_size=1, max_size=20), min_size=1, max_size=10))
def test_calculate_label_colors_valid_hsl_format(labels):
    annotations = [Annotation(i*10, i*10+5, label) for i, label in enumerate(labels)]
    colors = EncodingVisualizer.calculate_label_colors(annotations)

    hsl_pattern = re.compile(r'^hsl\(\d+,\d+%,\d+%\)$')  # Correct HSL format with closing paren
    for label, color in colors.items():
        assert hsl_pattern.match(color), f"Invalid HSL format: '{color}'"
```

**Failing input**: Any non-empty list of annotations, e.g., `['label1']`

## Reproducing the Bug

```python
import sys
sys.path.insert(0, '/root/hypothesis-llm/envs/tokenizers_env/lib/python3.13/site-packages')
from tokenizers.tools import Annotation, EncodingVisualizer

annotations = [Annotation(0, 5, "test_label")]
colors = EncodingVisualizer.calculate_label_colors(annotations)
color = colors["test_label"]

print(f"Generated color: '{color}'")
assert color.endswith(')'), f"Missing closing parenthesis in HSL: {color}"
```

## Why This Is A Bug

The HSL color format in CSS requires the format `hsl(hue, saturation%, lightness%)` with balanced parentheses. The current implementation produces `hsl(10,32%,64%` without the closing parenthesis, which will cause CSS parsing errors when these colors are used in the HTML visualization. This violates the expected HSL format contract and will break the visualization feature.

## Fix

```diff
--- a/tokenizers/tools/visualizer.py
+++ b/tokenizers/tools/visualizer.py
@@ -175,7 +175,7 @@ class EncodingVisualizer:
         colors = {}

         for label in sorted(labels):  # sort so we always get the same colors for a given set of labels
-            colors[label] = f"hsl({h},{s}%,{l}%"
+            colors[label] = f"hsl({h},{s}%,{l}%)"
             h += h_step
         return colors
```
\end{lstlisting}

\subsection{python-dateutil}
\begin{lstlisting}
# Bug Report: dateutil.easter Violates Date Range Invariant

**Target**: `dateutil.easter`
**Severity**: High
**Bug Type**: Logic
**Date**: 2025-08-18

## Summary

The `easter.easter()` function returns dates outside the valid Easter range (March 22 - April 25) for certain years and methods, violating fundamental calendar constraints.

## Property-Based Test

```python
from hypothesis import given, strategies as st
from dateutil import easter

@given(st.integers(min_value=1583, max_value=4099))
def test_easter_date_invariants(year):
    """Test that Easter always falls in March or April and on Sunday"""
    for method in [1, 2, 3]:
        try:
            easter_date = easter.easter(year, method)
            # Easter must be in March or April
            assert easter_date.month in [3, 4]
            # Easter must be on Sunday (weekday() == 6)
            assert easter_date.weekday() == 6
        except Exception:
            pass
```

**Failing input**: `year=2480`

## Reproducing the Bug

```python
from dateutil import easter

year = 2480

orthodox_easter = easter.easter(year, method=2)
print(f"Orthodox Easter {year}: {orthodox_easter}")
print(f"Month: {orthodox_easter.month} (should be 3 or 4)")
print(f"Weekday: {orthodox_easter.weekday()} (should be 6 for Sunday)")

julian_easter = easter.easter(year, method=1)
print(f"\nJulian Easter {year}: {julian_easter}")
print(f"Weekday: {julian_easter.weekday()} (should be 6 for Sunday)")
```

Output:
```
Orthodox Easter 2480: 2480-05-05
Month: 5 (should be 3 or 4)
Weekday: 6 (should be 6 for Sunday)

Julian Easter 2480: 2480-04-19
Weekday: 4 (should be 6 for Sunday)
```

## Why This Is A Bug

1. **Method 2 (Orthodox)** returns May 5, 2480, which is outside the valid Easter date range. Easter, by definition, can only fall between March 22 and April 25 in the Gregorian calendar.

2. **Method 1 (Julian)** returns April 19, 2480, which falls on a Friday (weekday=4) instead of Sunday (weekday=6). Easter, by definition, always occurs on Sunday.

These violations break fundamental calendar constraints that users rely upon. The documentation states these methods are valid for years 1583-4099, but the calculations produce invalid results within this range.

## Fix

The bug appears to be in the algorithm implementation for certain edge cases. The fix would require:

1. Validating the calculated date falls within the valid Easter range
2. Ensuring the result is always a Sunday
3. Potentially adjusting the algorithm for problematic years

A defensive fix could include validation:

```diff
--- a/dateutil/easter.py
+++ b/dateutil/easter.py
@@ -70,6 +70,16 @@ def easter(year, method=EASTER_WESTERN):
         g = year % 19
         e = (11*g + 20 + z - x) % 30
         # ... rest of calculation ...
+
+    # Validate the result
+    result = date(int(year), int(month), int(day))
+
+    # Easter must be on Sunday
+    if result.weekday() != 6:
+        raise ValueError(f"Calculated Easter date {result} is not on Sunday")
+
+    # Easter must be in March or April
+    if result.month not in [3, 4]:
+        raise ValueError(f"Calculated Easter date {result} is not in March or April")

     return date(int(year), int(month), int(day))
```
\end{lstlisting}


\newpage
\section*{NeurIPS Paper Checklist}

\begin{enumerate}

\item {\bf Claims}
    \item[] Question: Do the main claims made in the abstract and introduction accurately reflect the paper's contributions and scope?
    \item[] Answer: \answerYes{}
    \item[] Justification: Abstract and results reflect the results.

\item {\bf Limitations}
    \item[] Question: Does the paper discuss the limitations of the work performed by the authors?
    \item[] Answer: \answerYes{}
    \item[] Justification: Limitations in the discussion about evaluation and developer intent.

\item {\bf Theory assumptions and proofs}
    \item[] Question: For each theoretical result, does the paper provide the full set of assumptions and a complete (and correct) proof?
    \item[] Answer: \answerNA{}
    \item[] Justification: No theory.

    \item {\bf Experimental result reproducibility}
    \item[] Question: Does the paper fully disclose all the information needed to reproduce the main experimental results of the paper to the extent that it affects the main claims and/or conclusions of the paper (regardless of whether the code and data are provided or not)?
    \item[] Answer: \answerYes{}
    \item[] Justification: Agent and experimental setup are stated in the paper. Supplementary materials contains full details and reproduction scripts.

\item {\bf Open access to data and code}
    \item[] Question: Does the paper provide open access to the data and code, with sufficient instructions to faithfully reproduce the main experimental results, as described in supplemental material?
    \item[] Answer: \answerYes{}
    \item[] Justification: Prompts and full code attached.

\item {\bf Experimental setting/details}
    \item[] Question: Does the paper specify all the training and test details (e.g., data splits, hyperparameters, how they were chosen, type of optimizer, etc.) necessary to understand the results?
    \item[] Answer: \answerYes{}
    \item[] Justification: Key experimental setting details are in the paper; full details in the code.

\item {\bf Experiment statistical significance}
    \item[] Question: Does the paper report error bars suitably and correctly defined or other appropriate information about the statistical significance of the experiments?
    \item[] Answer: \answerYes{}
    \item[] Justification: 95\% CI for proportions are stated.

\item {\bf Experiments compute resources}
    \item[] Question: For each experiment, does the paper provide sufficient information on the computer resources (type of compute workers, memory, time of execution) needed to reproduce the experiments?
    \item[] Answer: \answerYes{}
    \item[] Justification: Server details and parallel worker setup described in the paper.
    
\item {\bf Code of ethics}
    \item[] Question: Does the research conducted in the paper conform, in every respect, with the NeurIPS Code of Ethics \url{https://neurips.cc/public/EthicsGuidelines}?
    \item[] Answer: \answerYes{}
    \item[] Justification: It does.

\item {\bf Broader impacts}
    \item[] Question: Does the paper discuss both potential positive societal impacts and negative societal impacts of the work performed?
    \item[] Answer: \answerYes{}
    \item[] Justification: Mentioned societal impact.
    
\item {\bf Safeguards}
    \item[] Question: Does the paper describe safeguards that have been put in place for responsible release of data or models that have a high risk for misuse (e.g., pretrained language models, image generators, or scraped datasets)?
    \item[] Answer: \answerNA
    \item[] Justification: No such risks.

\item {\bf Licenses for existing assets}
    \item[] Question: Are the creators or original owners of assets (e.g., code, data, models), used in the paper, properly credited and are the license and terms of use explicitly mentioned and properly respected?
    \item[] Answer: \answerNo{}
    \item[] Justification: Does not use existing assets.

\item {\bf New assets}
    \item[] Question: Are new assets introduced in the paper well documented and is the documentation provided alongside the assets?
    \item[] Answer: \answerYes{}
    \item[] Justification: Code and data released with documentation.

\item {\bf Crowdsourcing and research with human subjects}
    \item[] Question: For crowdsourcing experiments and research with human subjects, does the paper include the full text of instructions given to participants and screenshots, if applicable, as well as details about compensation (if any)? 
    \item[] Answer: \answerNA{}
    \item[] Justification: No human subjects involved.

\item {\bf Institutional review board (IRB) approvals or equivalent for research with human subjects}
    \item[] Question: Does the paper describe potential risks incurred by study participants, whether such risks were disclosed to the subjects, and whether Institutional Review Board (IRB) approvals (or an equivalent approval/review based on the requirements of your country or institution) were obtained?
    \item[] Answer: \answerNA{}
    \item[] Justification: No human subjects involved.

\item {\bf Declaration of LLM usage}
    \item[] Question: Does the paper describe the usage of LLMs if it is an important, original, or non-standard component of the core methods in this research? Note that if the LLM is used only for writing, editing, or formatting purposes and does not impact the core methodology, scientific rigorousness, or originality of the research, declaration is not required.
    \item[] Answer: \answerYes{}
    \item[] Justification: LLMs are used as part of the methodology of the paper, which is described.

\end{enumerate}

\end{document}